\documentclass[twocolumn,amsmath,amssymb]{revtex4-1}
\usepackage{amsmath}
\usepackage{graphicx}
\usepackage{dcolumn}
\usepackage{bm}
\usepackage{epsfig,color,xspace,multirow,xr,bbold}
\usepackage[all]{xy}
\usepackage{setspace}
\usepackage{url}

\newcommand{\qho}{{\rm QHO}}
\newcommand{\vdw}{{\rm vdW}}
\newcommand{\br}{{\bf r}}
\newcommand{\rd}{{\rm d}}
\newcommand{\rpq}{r_{pq}}
\newcommand{\ai}{{\rm ab initio}\xspace}

\begin{document}

\title{Toward transferable interatomic van der Waals interactions without
  electrons: The role of multipole electrostatics and many-body dispersion}
 
\author{Tristan Bereau}
\email{bereau@mpip-mainz.mpg.de}
\affiliation{Max-Planck-Institut f\"ur Polymerforschung, Ackermannweg 10, 55128
  Mainz, Germany}
\affiliation{Department of Chemistry,
  University of Basel, 4056 Basel, Switzerland} 
\author{O. Anatole von Lilienfeld}
\affiliation{Institute of Physical Chemistry, 
Department of Chemistry,
  University of Basel, 4056 Basel, Switzerland} 
\affiliation{Argonne Leadership Computing Facility, Argonne National
  Laboratory, Argonne, Illinois 60439, USA}
\date{\today}

\begin{abstract}
  We estimate polarizabilities of atoms in molecules without electron density,
  using a Voronoi tesselation approach instead of conventional density
  partitioning schemes.  The resulting atomic dispersion coefficients are
  calculated, as well as many-body dispersion effects on intermolecular
  potential energies.  We also estimate contributions from multipole
  electrostatics and compare them to dispersion.  We assess the performance of
  the resulting intermolecular interaction model from dispersion and
  electrostatics for more than 1,300 neutral and charged, small organic
  molecular dimers.  Applications to water clusters, the benzene crystal, the
  anti-cancer drug ellipticine---intercalated between two Watson-Crick DNA
  base pairs, as well as six macro-molecular host-guest complexes highlight
  the potential of this method and help to identify points of future
  improvement.  The mean absolute error made by the combination
  of static electrostatics with many-body dispersion reduces at larger
  distances, while it plateaus for two-body dispersion, in conflict with the
  common assumption that the simple $1/R^6$ correction will yield proper
  dissociative tails.  Overall, the method achieves an accuracy well within
  conventional molecular force fields while exhibiting a simple parametrization 
  protocol.
\end{abstract}

\maketitle

\section{Introduction}

Predicting potential energies from first principles is a long-standing
endeavor in physical chemistry.  Post-Hartree-Fock methods offer a great deal
of accuracy and transferability by solving Schr\"odinger's equation, though at
the expense of unfavorable scaling as a function of system
size.\cite{szabo1989modern} Density functional theory (DFT), on the other
hand, implies only cubic scaling through the use of electron density and
single-particle orbitals alone.\cite{parr1989density} DFT has proven
tremendously useful for the simulation of chemical systems and even for
obtaining thermodynamical properties through the use of {\em ab initio}
molecular dynamics.\cite{carparrinello,AIMD_PNAS_TUCKERMAN2005} Unfortunately,
common approximations to the employed exchange-correlation potential can
result in significant shortcomings, such as the inadequate---occasionally even
qualitatively wrong---predictions of the intermolecular forces that arise due
to noncovalent van-der-Waals (vdW) binding,\cite{Pulay-NoVdw} or a dramatic
overestimation of molecular polarizabilities.\cite{ChemistsGuidetoDFT} While
the latter can usually be cured through the use of hybrid density
functionals,\cite{FCACP_anatoleMP2013} the issue of vdW binding has motivated
the development of a wide variety of dispersion-corrected
methods,\cite{riley2010stabilization, french2010long, kannemann2010van} many
of which offer an atom-pairwise additive ad-hoc correction with the correct
dissociative polynomial behavior of London's dispersion formula, $C_6/R^6$,
referred to in the following as two-body dispersion
(TBD).\cite{Sprik-NoVdw,Grimme2004,Grimme2006} Even though the addition of
such atom pairwise corrections very much improve the accuracy of
DFT,\cite{GrimmeRxn2006} it has also been demonstrated that the interatomic
many-body vdW forces are not properly accounted for within LDA, GGAs,
meta-GGAs, and hybrid functionals.\cite{AlexMe3BodyDFT} Using the
Axilrod-Teller-Muto expression, the magnitude of interatomic three-body
dispersion forces has subsequently been found to be quite significant in many
systems, particularly when going beyond isolated molecular dimers toward
larger molecular assemblies or condensed-phase systems such as molecular
liquids or crystals.\cite{von2010two} To properly account for many-body
effects,\cite{linder2004many, renne1967microscopic} a many-body dispersion
(MBD) method has recently been introduced\cite{tkatchenko2012accurate} that
builds on the Tkatchenko-Scheffler (TS)\cite{tkatchenko2009accurate} TBD.
Within MBD, free-atom polarizabilities are coupled at short interatomic
distances by means of partitioning the atom around its closest neighbors
according to its electron density.  The MBD energy up to infinite order (i.e.,
$R^{-n}$, $n = \{6, 9, \ldots\}$) is then obtained by diagonalizing the
Hamiltonian corresponding to a system of coupled fluctuating harmonic
dipoles,\cite{donchev2006many} thereby coupling the polarizabilities at long
range as well.  The importance of MBD has been demonstrated for a large
variety of molecular assemblies and systems.\cite{tkatchenko2012accurate,
  distasiomany, distasio2012collective, marom2013many}

In this work, we probe the ability to represent dispersion interactions using
the TBD and MBD methods \emph{without} an underlying electron density.  Though
initially developed to correct DFT calculations, the model provides a sound
and efficient way to compute two-body and many-body dispersion with a minimal
amount of free parameters and assumptions.  This strategy follows a number of
extensive efforts toward modeling largely classical potentials derived from
rigorous quantum-mechanics, such as SIBFA\cite{piquemal2007toward,
  piquemal2007anisotropic} and EFP,\cite{gordon2013accurate} that partition
the interaction energy into specific terms, e.g., static electrostatics,
polarization, repulsion, dispersion, charge transfer.  As a promising
compromise, the AMOEBA force field\cite{ponder2010current} provides a number
of refinements beyond standard atomistic models (e.g., multipole
electrostatics, polarization) that leads the way toward next-generation force
fields.\cite{antila2013polarizable} Far from competing with these established
methods, the present work explores the prospects of an ``electron-free'' yet
physics-based approach to dispersion interactions.  The present work only
considers dispersion alongside static multipole
electrostatics,\cite{kramer2013deriving} and thus lacks critical terms toward
a rigorous intermolecular potential (more below).  To the best of our
knowledge, such a systematic model of MBD interactions has not yet been
applied to compute intermolecular energies without any underlying electron
density---all abovementioned force fields and potentials rely instead on
pairwise interactions.  Alternatively, the TBD approach we use here, though
based on a standard $R^{-6}$ functional form, adapts the associated $C_6$
coefficients to the local chemistry (e.g., hybridization) and geometry of the
system.  The common usage of $R^{-6}$ functions and the present TBD method's
limited computational overhead makes it relevant for force-field-based
molecular modeling.

In Sec.~\ref{sec:model}, we introduce the approximations and changes necessary
to compute MBD and TBD energies without electron density, i.e. from Voronoi
partitioning only.  Details for calculating the MTP electrostatics are also
provided.  Next, we optimize the free parameters to best reproduce numbers for
a training set of experimental molecular polarizabilities, as well as a small
set of intermolecular energies of gas-phase dimers (Sec.~\ref{sec:param}).  We
then report the method's performance for a large variety of systems:
Sec.~\ref{sec:results} reports on intermolecular energies for a diverse set of
one thousand hydrogen-bonded and dispersion-bonded complexes, neutral and
charged amino-acid side chains, ionic groups, organic halides, halohydrides,
and halogen molecules.  Thereafter, we assess the method in more detail for a
select set of interesting systems.  Specifically, we discuss potential
energies of binding for the water dimer, thirty-eight water clusters with 2 to
10 monomers, the benzene crystal, and seven supramolecular host-guest
complexes including the aromatic ellipticine drug and DNA base-pair dimer
complex.

\section{Methods}
\label{sec:model}

\subsection{Effective atomic polarizabilities}

The frequency-dependent polarizability of atom $p$ can be expressed as a
truncated Pad\'e series
\begin{equation}
  \label{eq:ts}
  \alpha_p (i\omega) \approx \frac{\alpha_p^0[n(\br)]}{1 + \{
    \omega/\omega_p[n(\br)]\}^2},
\end{equation}
where $\alpha_p^0[n(\br)]$ is the static polarizability, $\omega_p[n(\br)]$ is
the corresponding characteristic excitation frequency, and $n(\br)$ is the
electron density.  The two quantities $\alpha_p$ and $\omega_p$ in
Eq.~(\ref{eq:ts}) are linked to the leading pairwise dispersion coefficient
term $C_6$ for atoms $p$ and $q$ via the Casimir-Polder integral
\cite{casimir1948influence}
\begin{equation}
  \label{eq:casimir}
  C_{6pq} = \frac 3\pi \int_0^\infty \rd\omega \alpha_p (i\omega)
  \alpha_q (i\omega).
\end{equation}
For $p=q$, this yields $\omega_p = 4C_{6pp}/3(\alpha_p^0)^2$.  Tkatchenko and
Scheffler (TS) proposed to compute $\alpha^0_p$ of an atom in a molecule via
the ratio of its effective and free volumes, $\alpha^0_p = \alpha^{\rm free}_p
(V_p^{\rm eff}/V_p^{\rm free})$, as obtained through Hirshfeld partitioning of
electron densities,
\begin{equation}
  \frac{V_p^{\rm eff}}{V_p^{\rm free}} = \frac{\int {\rm d}{\bf r} r^3
    w_p({\bf r})n({\bf r})} {\int {\rm d}{\bf r} r^3 n_p^{\rm free}({\bf
      r})}.\label{eq:hirshfeld} 
\end{equation}
Here, $n_p^{\rm free}({\bf r})$ is the electron density of the free atom $p$
and $w_p({\bf r})$ weighs the contribution of the free atom $p$ with respect
to all free atoms at position ${\bf r}$.\cite{tkatchenko2009accurate}
Accurate free-atom reference values of $\alpha_p$ and $C_{6pp}$ can be found
in a study from Chu and Dalgarno.\cite{chu2004linear}

In this work we relax the requirement for an electron density $n(\br)$.
Instead, the free-atom density is represented by a Gaussian function centered
at the nucleus, and the free-atom radius, $R_p^\vdw$,\cite{von2010two} is used
as its width.  We also replace the total electron density in the numerator by
this free-atom density.  As a proxy for the weight $w_p({\bf r})$, we
partition space into two regions:
\begin{equation}
  w_p({\bf r}) \approx \left\{
  \begin{array}{lc}
    1, & {\bf r} \in \mathcal{R}_p\\
    \exp(-d({\bf r},{\bf r}_p)/(d_w R_p^\vdw)), & {\rm otherwise},
  \end{array}
  \right.
\end{equation}
where ${\bf r}_p$ is the position of atom $p$, $d({\bf r}, {\bf r}_p) = |{\bf
  r} - {\bf r}_p|$ is the Euclidean distance between points ${\bf r}$ and
${\bf r}_p$, and $d_w$ is an atom-type independent, free parameter that scales
the exponential decay.  The partitioning between the region belonging to atom
$p$, $\mathcal{R}_p$, and the rest relies on a Voronoi-type tessellation
scheme \cite{okabe2009spatial} to determine which atom is closest to any
position ${\bf r}$
\begin{equation}
  \label{eq:voronoi}
  \mathcal{R}_p = \{{\bf r} \in \mathbb{R}^3 \;|\; d({\bf r}, {\bf r}_p) \leq
  d({\bf r}, {\bf r}_j)\; \text{for all}\; j \neq p\}.
\end{equation}
In practice, $\mathcal{R}_p$ is evaluated from a set of points ${\bf r}$
located around atom $p$.  While the Voronoi scheme systematically aims at the
midplane between two atoms, the asymmetry between different chemical elements
is encoded in $w_p({\bf r})$ via the free-atom radii $R_p^\vdw$.  To evaluate
Eq.~(\ref{eq:hirshfeld}), we perform a discrete integration over a cubic grid
of size 20 Bohr$^3$ with step sizes of 1.0 Bohr around each atom $p$.

We further note that this partitioning does not allow to distinguish energetic
differences between different charged states of a molecule from dispersion.
Any effect due to a net charge is hereby only imprinted in the static
electrostatics. 

\subsection{Many-body Dispersion (MBD) energy}
\label{sec:mbdenergy}

Originally, the MBD scheme prescribed a prior long-range coupling of the
atomic polarizabilities.\cite{tkatchenko2012accurate} We note that more
recently, it has been shown that the coupled quantum harmonic oscillator (QHO)
model already includes long-range electrodynamic
screening.\cite{ambrosetti2014long} In that procedure, short-ranged screening
effects are excluded from the MBD equation to avoid double counting of the DFT
correlation energy and the polarizability screening.  We hereby omit this
separation, since our scheme does not entail any correlation, and work instead
with the full interaction tensor.

The MBD energy is computed from the coupled fluctuating dipole model for a
collection of coupled isotropic QHOs, each representing an atom in the
system.\cite{donchev2006many, tkatchenko2012accurate} We rely on the
abovementioned set of effective atomic polarizabilities, $\alpha_p^0$, and
their corresponding characteristic frequencies, $\omega_p^0$.  The energy is
obtained by diagonalizing the $3N \times 3N$ matrix
\begin{equation}
  \label{eq:mbdeig}
  C_{pq}^\qho = (\omega_p)^2 \delta_{pq} + (1-\delta_{pq}) \omega_p\omega_q
  \sqrt{\alpha_p \alpha_q} \mathcal{T}_{pq},
\end{equation}
where $\mathcal{T}_{pq} = \nabla_{\br_p} \otimes \nabla_{\br_q} W(\rpq)$ is
the dipole interaction tensor, $\br_p$ and $\br_q$ are the atoms' positions,
and $\delta$ is Kronecker's delta.  Here, the dipole interaction tensor relies
on the modified Coulomb potential
\begin{equation}
  \label{eq:modcoul}
  W(\rpq) = \frac{1 - \exp\left[ -\left(\frac{\rpq}{R_{pq}^\vdw}\right)^\beta
      \right]}{\rpq} ,
\end{equation}
where $\rpq = |\br_p - \br_q|$, $\beta$ is a range-separation parameter and
$R_{pq}^\vdw$ is proportional to the sum of the vdW radii for a pair of atoms
$p$ and $q$: $R_{pq}^\vdw = \gamma_{\rm MBD} (R_p^\vdw + R_q^\vdw)$, where
$\gamma_{\rm MBD}$ is an atom-type independent, free parameter.  The vdW
radius of an atom is scaled according to the Tkatchenko-Scheffler scheme,
$R_p^\vdw = (\alpha_p/\alpha_p^{\rm free})^{1/3} R_p^{\vdw,{\rm free}}$.  The
free-atom vdW radii are formally obtained by measuring electron density
contours for rare-gas atoms and extending the result to other atoms of the
same row.\cite{tkatchenko2009accurate} All vdW radii used in the present work
are reported in Ref.~\citenum{von2010two}, except for iodine, which we
determined as $4.39$ Bohr.  The modified dipole interaction tensor for
cartesian components $a$ and $b$ between atoms $p$ and $q$ yields
\begin{widetext}
\begin{eqnarray}
  \mathcal{T}_{pq}^{\prime ab} &=& -\frac{3 \rpq^a \rpq^b - \rpq^2
    \delta_{ab}}{\rpq^5} 
  \left(  
    1-\exp\left[ -\left(\frac{\rpq}{R_{pq}^\vdw}\right)^\beta \right]
    -\beta\left(\frac{\rpq}{R_{pq}^\vdw}\right)^\beta \exp \left[
      -\left(\frac{\rpq}{R_{pq}^\vdw}\right)^\beta \right] 
  \right) \\*
  \nonumber
  &&+ \left( \beta \left(\frac{\rpq}{R_{pq}^\vdw}\right)^\beta + 1 -\beta
  \right) \beta \left(\frac{\rpq}{R_{pq}^\vdw}\right)^\beta
  \frac{\rpq^a\rpq^b}{\rpq^5} \exp\left[
    -\left(\frac{\rpq}{R_{pq}^\vdw}\right)^\beta \right]. 
\end{eqnarray}
\end{widetext}

The interaction energy is given by the difference of the coupled and uncoupled
systems of QHOs
\begin{equation}
  E_{\rm MBD} = \frac 12 \sum_{i=1}^{3N} \sqrt{\lambda_i} - \frac 32
  \sum_{p=1}^N \omega_p,
\end{equation}
where $\{\lambda_i\}$ are the eigenvalues of the matrix $C^\qho$
(Eq.~\ref{eq:mbdeig}).  Within the approximations of the model, the MBD energy
includes \emph{all} contributions of the dispersion interaction, including the
pairwise component.

\subsection{Molecular polarizability}

The MBD eigenvalue problem (Eq.~\ref{eq:mbdeig}) has been shown equivalent to
the calculation of molecular polarizability from a dipole-dipole electric
field coupling equation.\cite{applequist1972atom} As described by
Applequist,\cite{applequist2003polarizability} the molecular polarizability
can be obtained from a combination of the eigenvalues and eigenvectors of
Eq.~\ref{eq:mbdeig}, the norm of eigenvector $i$ being subsequently scaled by
$\alpha_i^{-1/2}\omega_i^{-1}$.  Though all atomic polarizabilities and
characteristic frequencies considered up to now were purely isotropic, we have
found strongly enhanced molecular polarizability anisotropies when using
\emph{anisotropic} coefficients for the eigenvector scaling.  To do so, we
replace the $r^3$ factor in Eq.~\ref{eq:hirshfeld} by $({\bf r}\cdot{\bf
  e}_{\alpha})^2r$, where ${\bf e}_\alpha$ is a cartesian unit vector along
$\alpha \in \{x,y,z\}$.

\subsection{Two-body Dispersion (TBD) energy}
While the reward for using MBD is evident, inclusion of TBD in the discussion
also has its merits.  The TBD amounts to (i) a commonly used DFT correction,
(ii) the dissociative Lennard-Jones tail employed for intermolecular
non-covalent energy contributions in many force-fields, (iii) an interesting
comparison of two-body versus many-body effects, and (iv) a straightforward
way to evaluate analytic force contributions.  As such, our TBD implementation
suggests a simple recipe to approximately include the dynamic, through atomic
Voronoi partitioning re-evaluated ``on-the-fly,'' dispersion coefficients in
MD packages, as it only requires the estimation of effective polarizabilities
to determine the $C_6$ coefficients.  Therefore, and following the TS
prescription,\cite{tkatchenko2009accurate} we also extend the discussion to
include dispersion energies arising from exclusively pairwise dipole-dipole
interactions,
\begin{equation}
  E_{\rm TBD} = - \frac 12 \sum_{p,q} f_{\rm damp}(r_{pq},R_p^\vdw,R_q^\vdw)
  C_{6pq} r^{-6}_{pq}, 
\end{equation}
where
\begin{equation}
  C_{6pq} = \frac{2 C_{6pp}C_{6qq}}{\frac{\alpha_q}{\alpha_p}C_{6pp} +
    \frac{\alpha_p}{\alpha_q}C_{6qq}},
\end{equation}
the $C_{6}$ and $\alpha$ parameters are computed from the frequency-dependent
polarizability $\alpha(i\omega)$, and we use a Fermi-type damping function
\begin{eqnarray}
  f_{\rm damp} &(&r_{pq},R_p^\vdw,R_q^\vdw) = \notag\\
  && \frac 1{1 + \exp\left[ -d\left(
      \frac{r_{pq}}{s_R (R_p^\vdw + R_q^\vdw)} - 1\right)\right]},
\label{eq:damp}
\end{eqnarray}
where $d$ and $s_R$ are unitless, free parameters obtained through
optimization (see below).

\subsection{Multipole electrostatics}

Electrostatic interactions were computed using a static multipolar expansion.
The electrostatic potential (ESP), $\Phi({\bf r}) = \int \rd {\bf r}' n({\bf
  r'})/|{\bf r} - {\bf r'}|$, of a charge density $n$ at position ${\bf r}$ is
expanded in a Taylor series of $1/R \equiv 1/|{\bf r} - {\bf r'}|$, providing
the following MTP expansion for $\Phi$ in Cartesian coordinates
\begin{equation}
  \label{eq:mtpfit}
  \Phi({\bf r}) = \frac qR + \frac{\mu_\alpha R_\alpha}{R^3} +
  \frac 13 \Theta_{\alpha\beta} \frac{3R_\alpha R_\beta -
    R^2\delta_{\alpha\beta}} {R^5} + \ldots,
\end{equation}
where $q$ is a partial charge, $\mu_\alpha$ is the component $\alpha$ of the
dipole moment ${\bf \mu}$, $\Theta_{\alpha\beta}$ is the component
$\alpha\beta$ of the second-rank quadrupole moment tensor $\Theta$, and
summation over repeating indices is implied.  Analogous to Coulomb
interactions, the MTP interaction is determined by the work done on an MTP
$Q_{l\kappa}$ (i.e., order $l$ and index $\kappa$) brought from infinity to a
point $r$ in a region populated by the potential $\Phi$, $E_{\rm MTP} = \Phi
Q_{l\kappa}$.\cite{stone2013theory} For this study, MTP coefficients up to
quadrupoles have been placed on each atomic site.  All multipole interactions
were computed in CHARMM \cite{brooks2009charmm} using the MTPL
module.\cite{bereau2013leveraging, bereau2013scoring}

\section{Results}
\label{sec:results}

\section{Parameter optimization}
\label{sec:param}

\subsection{Multipole coefficients}

MTP coefficients were obtained by means of a fit to the \ai ESP of each
compound, using second-order M{\o}ller Plesset and an aug-cc-pVDZ basis set
(except for the supramolecular host-guest complexes, see below).  For each
molecule, we computed the ab initio ESP of the conformation provided by the
reference geometry, without further optimization.  MTP parameters were then
fitted to best reproduce the ESP in the close vicinity of the molecule (see
Eq.~\ref{eq:mtpfit}).  To avoid fitting artifacts due to insufficient
sampling of buried atoms, we constrained each monopole to deviate at most by
an amount $0.1~e$ from the corresponding monopole derived from the generalized
distributed multipole analysis (GDMA).\cite{stone2013theory}  More details on
the ESP-based fitting protocol, as well as scripts to reproduce the present
results, can be found in Kramer et al.\cite{kramer2013deriving}

\subsection{Polarizabilities}

\begin{figure}[htbp]
  \begin{center}
    \includegraphics[width=\linewidth]{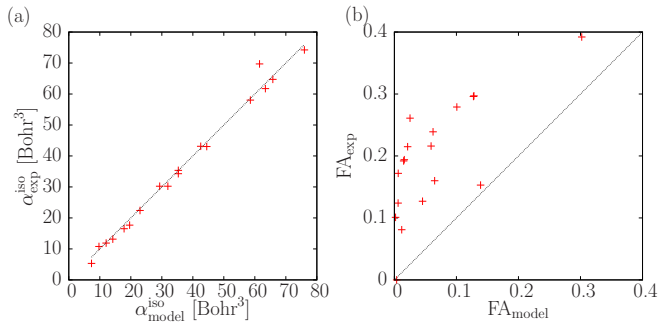}
    \caption{Correlation plot between (a) isotropic and (b) fractional
      anisotropies of molecular polarizabilities predicted from the current
      implementation and experimental values for the set of 18 compounds
      proposed in Ref.~\citenum{thole1981molecular}.}
    \label{fig:alphas}
  \end{center}
\end{figure}

The estimation of molecular polarizabilities based on Voronoi partitioning
requires the determination of three free parameters: ($i$) $d_w$, which
controls the decay of the weight $w_p({\bf r})$ in the polarizability
calculation (Eq.~\ref{eq:hirshfeld}), ($ii$) $\beta$ from the modified Coulomb
potential (Eq.~\ref{eq:modcoul}), and ($iii$) the radius prefactor
$\gamma_{\rm MBD}$ (see Sec.~\ref{sec:mbdenergy}).  We have found, however,
that the latter two affect molecular polarizabilities negligibly, within a
reasonable range---they will be parametrized in the next Section.  We optimize
$d_w$ globally, i.e. it is not atom-type dependent, to best reproduce a set of
18 experimentally-determined molecular polarizabilities benchmarked in
Refs.~\citenum{thole1981molecular}, \citenum{tkatchenko2012accurate}, and
\citenum{distasiomany}.  The parameter was systematically varied to minimize
the mean-absolute relative error (MARE: $\sum_i |\alpha_i^{\rm exp} -
\alpha_i^{\rm model}|/\alpha_i^{\rm exp}$) of the isotropic molecular
polarizabilities.  For $d_w = 3.8$ Bohr$^{-1}$, we find minimal MARE values of
6\% for the isotropic polarizabilities.  This outcome compares favorably with
the 9.1\% MARE found by DiStasio et al.\cite{distasiomany} Fractional
anisotropies (FA: $1/4\{[(\alpha_{xx} - \alpha_{yy})^2 + (\alpha_{xx} -
\alpha_{zz})^2 + (\alpha_{yy} - \alpha_{zz})^2]/(\alpha_{xx}^2 + \alpha_{yy}^2
+ \alpha_{zz}^2)\}^{1/2}$) are not as well accounted for: 68\%, while they
found 34\%.  Fig.~\ref{fig:alphas} shows the scatter plot of (a) the isotropic
molecular polarizabilities and (b) fractional anisotropies predicted from the
best-performing parameters versus experimental values used for fitting.  We
find decent correlation for the fractional anisotropies against the
experimental values, though our model systematically underestimates them.
This systematic effect hints at an imbalance in the dipole interaction tensor,
though none of its free parameters ($\gamma_{\rm MBD}$ and $\beta$) affects
significantly the fractional anisotropies.  The more pronounced effect of
$d_w$ suggests the role of the quality of the atomic polarizabilities
themselves.  A simple Voronoi partitioning scheme is likely not sufficient to
appropriately describe the anisotropy of molecular polarizabilities.

\subsection{MBD interaction energies}

To compute relaxed geometries within MBD, two additional parameters required
optimization, the range-separation parameter, $\beta$, of the modified Coulomb
potential (Eq.~\ref{eq:modcoul}), and the radius prefactor $\gamma_{\rm MBD}$
that scales $R_{pq}^{\vdw}$ (see Sec.~\ref{sec:mbdenergy}).  These parameters
were systematically varied by adding up MBD and MTP energies to best reproduce
the set of intermolecular dimer energies from the S22 dataset
\cite{jurevcka2006benchmark}---a benchmark database with 22 high-level quantum
chemistry (CCSD(T)) interaction energies for dimers of small molecules, DNA
base pairs, and amino acid pairs in their minimal potential energy geometry.
In this regard, the MTP intermolecular energies are assumed to be correct and
subtracted from the reference energies, such that we optimize the free
parameters to minimize the quantity $\sum_i |E_i^{\rm MBD} - (E_i^{\rm ref} -
E_i^{\rm MTP})|$.  The values $\beta = 1.10$ and $\gamma_{\rm MBD} = 1.85$
yield a minimal mean-absolute error (MAE: $\sum_i |E_i^{\rm ref} - E_i^{\rm
  model}|$) of 1.7~kcal/mol.  It is noteworthy that a number of common DFT
functionals were found to perform at similar level or
worse.\cite{DFTdispersionlessS22} Combination of MBD with PBE0 has been shown
to yield an error of only 0.3~kcal/mol.\cite{tkatchenko2012accurate} Our
results for S22 also compare well to energies predicted by a number of
standard force fields, Amber (2.1~kcal/mol), OPLS-AA (2.0~kcal/mol), and
MMFF94s (1.6~kcal/mol).\cite{paton2009hydrogen}

\begin{figure*}[htbp]
  \begin{center}
    \includegraphics[width=0.9\linewidth]{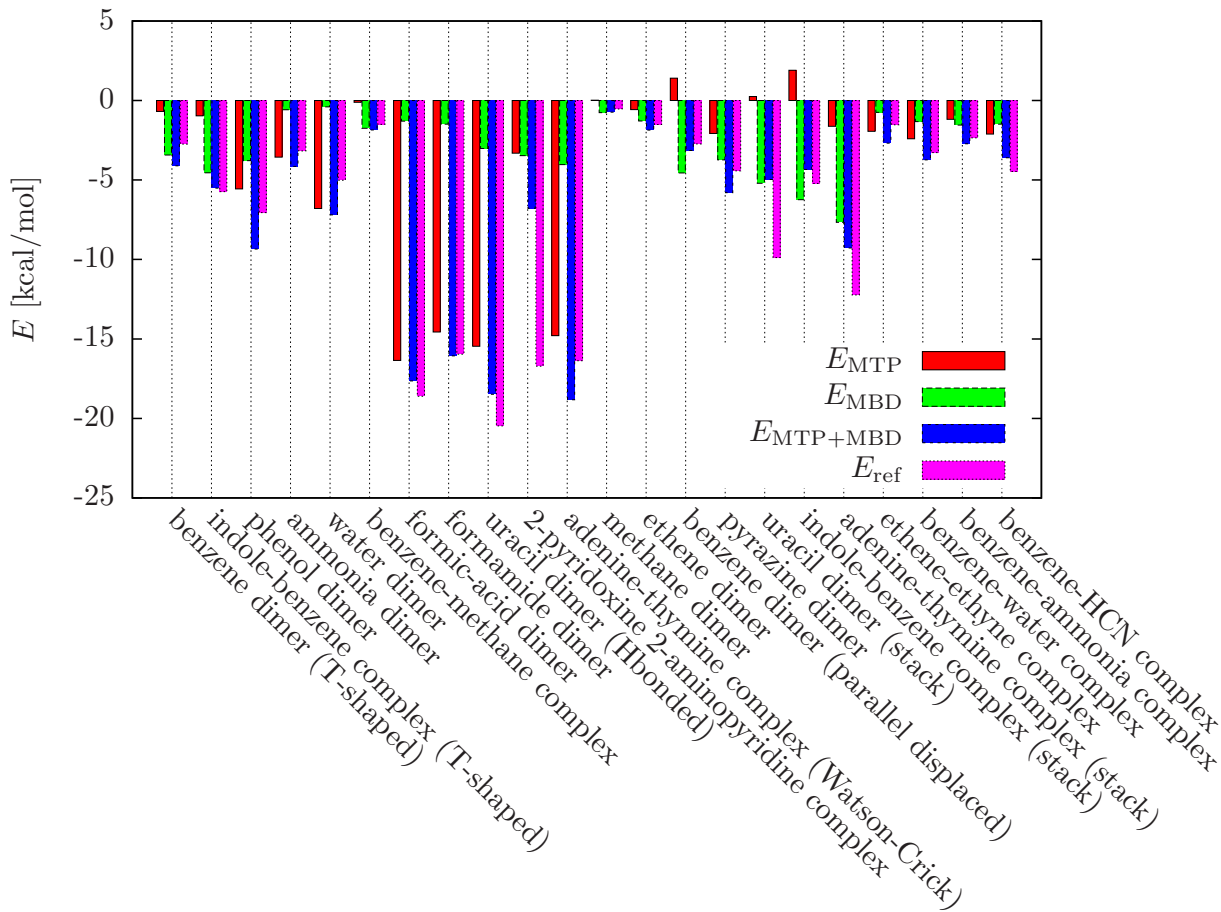}
    \caption{Intermolecular energy decomposition of the S22 dataset. MTP, MBD,
      MTP+MBD, and $E_{\rm ref}$ stand for multipole electrostatic, many-body
      dispersion, and reference energy (CCSD(T))
      contributions.\cite{jurevcka2006benchmark} This dataset has been used to
      optimize the MBD parameters in our model.}
    \label{fig:s22decomp}
  \end{center}
\end{figure*}

To analyze the performance of our MBD+MTP model, we display in
Fig.~\ref{fig:s22decomp} the energy decomposition between MTP and MBD together
with the reference energies for each compound in the S22 dataset.  As one
would expect, for the hydrogen-bonded dimers, the MTP contributions dominate,
with the MBD contribution typically improving the overall prediction. The most
dramatic failure for this class of dimers is found in the case of the
2-pyridoxine 2-aminopyridine complex, where MBD+MTP underestimates the actual
interaction energy by $\sim$10 kcal/mol.  At this point, we also stress the
lack of higher multipoles (i.e., beyond dipole) in the MBD method (e.g.,
$C_8$, $C_{10}$ coefficients). Jones et al.~have recently summarized the types
of dispersion interactions that arise from the different induced
multipoles.\cite{jones2013quantum} This error might also originate from the
lack of {\em induced} multipole moment contributions, which are not accounted
for within our model.  Another possible source of errors could be due to the
fact that in this work, we rely on the MBD energy alone to fully reproduce all
dispersion at short as well as long-range interatomic distances.  A
$\beta$-value smaller than 2 will yield a finite amount of dispersion energy
in the zero-distance limit.  We note that in contrast the augmentation of DFT
functionals with dispersion corrections typically implies $\beta \ge 2$, i.e.,
$W(r_{pq} \rightarrow 0) \rightarrow 0$, due to some amount of dispersion
energy already accounted for through the short range correlation in the
functional.\cite{tkatchenko2012accurate} Clearly, also capturing short-range
correlation effects still leaves room for further improvement.  In the case of
the dimers where the interaction energy is dominated by dispersion, the
combination of MTP and MBD predicts interaction energies that compare very
favorably with the reference numbers.  It is particularly encouraging that in
several cases an overestimation due to MBD is offset by a repulsive MTP
contributions resulting in an accurate overall interaction energy, e.g.,
indole-benzene complex (stack).  We also note that the entire MBD contribution
is attractive in nature for all the instances in the S22.

\subsection{TBD interaction energies}

In the same spirit as for MBD, TBD energies are compared to the reference
energies, subtracted by the MTP contribution.  Errors in TBD energies in the
S22 relaxed geometries were minimized through tuning of the dimensionless free
parameters $d$ and $s_R$ (Eq.~\ref{eq:damp}).  While $d=20.0$ is a
commonly-used value,\cite{tkatchenko2009accurate, marom2011dispersion,
  buvcko2013tkatchenko} we have found $d=11.0$ to perform best.  As for $s_R$,
a value of 2.20 proved to be best.  Overall, the error converges to a MAE of
2.0~kcal/mol, which also compares extremely favorably against other DFT
functionals without dispersion correction and standard force fields.

\subsection{$C_6$ coefficients from Voronoi partitioning}

\begin{figure}[htbp]
  \begin{center}
    \includegraphics[width=0.9\linewidth]{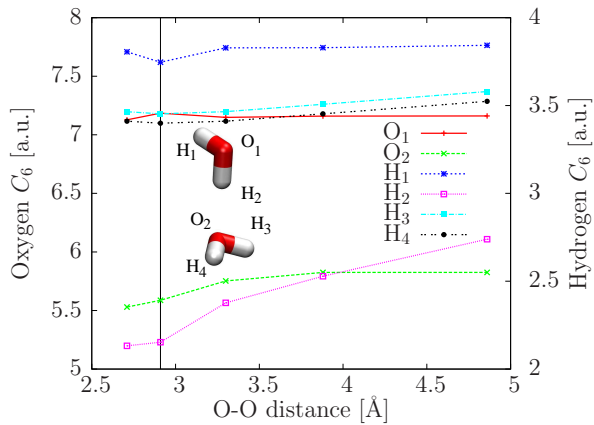} 
    \caption{$C_6$ coefficients obtained from Voronoi partitioning
      (Eq.~\ref{eq:casimir}) for all hydrogen and oxygen atoms in the water
      dimer as a function of the oxygen-oxygen distance.  $C_6$ coefficients
      in units Hartree Bohr$^6$.}
    \label{fig:c6}
  \end{center}
\end{figure}

To assess the performance of the Voronoi partitioning scheme for estimating
$C_6$ coefficients of atoms in molecules, we compute the $C_6$ coefficients of
the water dimer in the S22 dataset as a function of oxygen-oxygen distance.
Specifically, the water dimer exhibits $C_6$ values of $\approx 3$ and
$7$~Hartree Bohr$^6$ for the hydrogen and oxygen atoms, respectively, as shown
in Fig.~\ref{fig:c6}.  We find slightly large and low $C_6$ values for
hydrogen and oxygen atoms, respectively, compared to higher-level
calculations.\cite{wu2002empirical, tkatchenko2009accurate, von2010two} We
suggest that our simple Voronoi partitioning scheme does not sufficiently
account for the difference in electronegativity of the two atoms, in spite of
the partitioning's dependence to the free-atom's van der Waals radii.  To
further probe the $C_6$ coefficients obtained from Voronoi partitioning, we
computed their average values for each chemical element across the S22 dimers.
We find the following: C ($15.3 \pm 0.4$), O ($7.1 \pm 0.9$), N ($8.4 \pm
1.3$), and H ($3.1 \pm 0.4$), all expressed in Hartree Bohr$^6$.  Reference
values are in the ranges: C ($24-33$), O ($12-15$), N ($17-20$), and H
($2-3$), also in atomic units.\cite{tkatchenko2009accurate} Still, the
geometry-dependent variation of the $C_6$ coefficients is properly accounted
for and illustrates the coupling between the two molecules.

\subsection{Molecular dimers}

In Tab.~\ref{tab:mae}, we report the MAE of potential energies of interaction
for a total of over 1,300 geometries of small molecular dimers drawn from a
variety of databases designed to probe non-covalent interactions (extracted
from Ref.~\citenum{vrezavc2008quantum}).  These complexes include the
S22;\cite{jurevcka2006benchmark} its extension to nonequilibrium geometries,
S22x5;\cite{grafova2010comparative} S66---a larger coverage of interactions
found in organic molecules and biomolecules---and its extension to
nonequilibrium geometries, S66x8;\cite{rezac2011s66} hydrogen bonds found in
ionic groups interacting with neutral compounds (IonicHb)---constructed in
analogy to S66x8;\cite{rezac2011advanced} a set of 24 pairs of representative
amino acid side chain interactions (SCAI);\cite{berka2009representative} a set
of 40 complexes of organic halides, halohydrides, and halogen molecules at
equilibrium (X40) and nonequilibrium (X40x10)
geometries.\cite{rezac2012benchmark} Results are shown in Tab.~\ref{tab:mae}
for the respective MTP, TBD, and MBD contributions.  We remind the reader that
only S22 was used to fit the free parameters of the method.  As one would
expect, the sum of dispersion and electrostatics (referred to as ``MBD+MTP'')
performs on average better than any individual components, except in a few
isolated cases (i.e., see components of S66, S66x8, and IonicHb).  In the case
of S22, MBD+MTP performs on par with standard force fields for the SCAI
database: OPLS (2.1~kcal/mol) and Amber FF03
(2.2~kcal/mol).\cite{berka2009representative} The improvement for going from
TBD to MBD is largest for the SCAI-database, possibly due to the aromaticity
of a number of amino-acid side chains.

\begin{figure}[htbp]
  \begin{center}
    \includegraphics[width=0.8\linewidth]{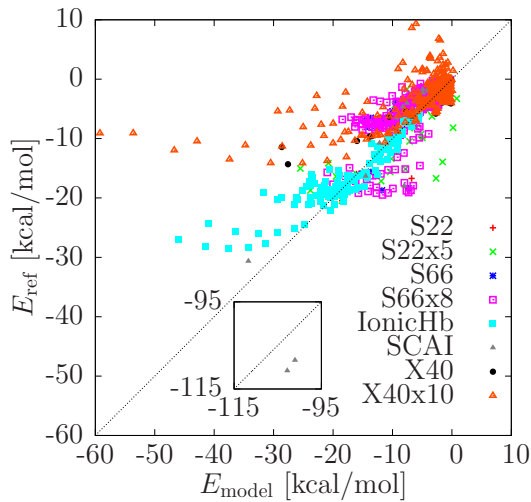}
    \caption{Correlation plot between energies predicted from the current
      implementation, $E_{\rm model}$ = MBD+MTP, and reference energies,
      $E_{\rm ref}$ for each database (Tab.~\ref{tab:mae}). The inset
      corresponds to two charged-charged side-chain interactions of the SCAI
      database.  Strong outliers from X40x10 exert overstabilizing MTP
      interactions (see main text).}
    \label{fig:corr}
  \end{center}
\end{figure}

\begin{figure}[htbp]
  \begin{center}
    \includegraphics[width=\linewidth]{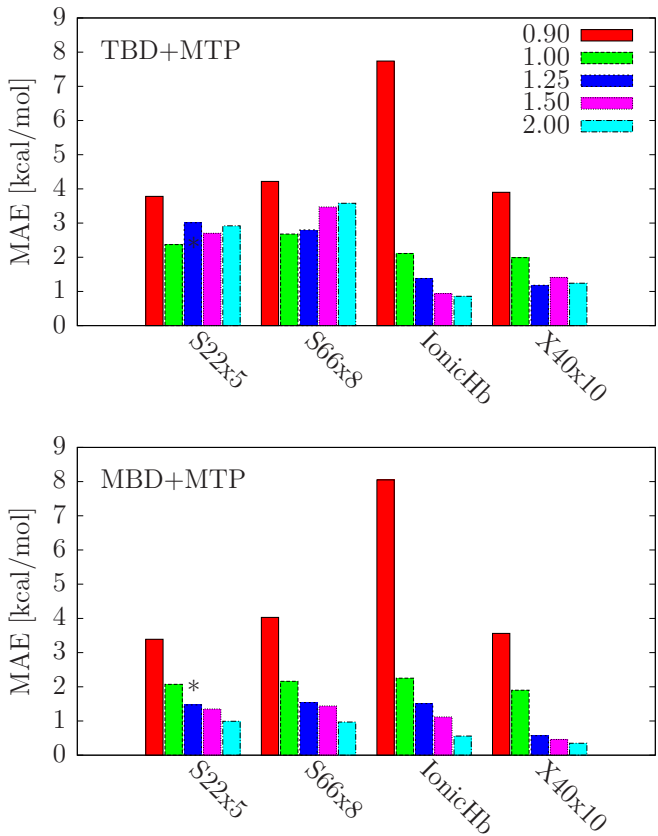}
    \caption{MAE as a function of distance relative to the equilibrium
      geometry for the databases S22x5, S66x8, IonicHb, and X40x10 for both
      TBD+MR (top) and MBD+MR (bottom).  The intermolecular distance at
      equilibrium geometry is scaled by a factor (i.e., 1.00 corresponds to
      the equilibrium geometry).  Note that the S22x5 at 1.25 is referring to
      a an energy that was in fact calculated using a factor of 1.20 instead
      (denoted by asterisks above the corresponding histograms).  }
    \label{fig:dist}
  \end{center}
\end{figure}

A scatter plot between all 1,300 intermolecular energies predicted from the
current implementation and reference energies (most of which are CCSD(T)) is
shown in Fig.~\ref{fig:corr}.  Clearly, the X40 and X40x10 data sets represent
the most difficult challenges for MBD+MTP. The MBD+MTP severely overestimates
the binding of HCl-methylamine, HF-methylamine, HF-methanol,
trichloromethanol-water, and trifluoromethanol-water---specifically for
distances shorter than the equilibrium geometry, where repulsive interactions
are critically lacking from our model (see also Fig.~\ref{fig:dist}). For
those, the MTP energy alone is significantly lower than the reference energy.
Similar overestimations are also encountered in the case of the larger
supramolecular guest-host systems discussed below, though the role of MTP
there is not as clear.

\begin{table}[htbp]
  \begin{center}
    \begin{tabular*}
      {\linewidth}{@{\extracolsep{\fill}} l|ccccc}
      & MTP & TBD & MBD & TBD+MTP & MBD+MTP  \\ 
      \hline
      S22     &  3.46 &   4.71 &   4.87 &   2.04 &   1.67\\
      S22x5   &  2.63 &   3.88 &   3.25 &   2.94 &   1.90\\
      S66     &  2.43 &   3.91 &   3.42 &   3.19 &   2.72\\
      S66x8   &  2.05 &   3.72 &   2.81 &   3.12 &   2.08\\
      IonicHb &  2.53 &  15.07 &  15.13 &   2.52 &   2.62\\
      SCAI    &  3.32 &  14.76 &  13.06 &   5.83 &   2.88\\
      X40     &  2.71 &   2.67 &   2.56 &   2.11 &   2.00\\
      X40x10  &  2.85 &   2.71 &   2.26 &   3.22 &   2.83\\
      \hline
      average &  2.75 &   6.43 &   5.92 &   3.12 &   2.34\\
    \end{tabular*}
    \caption{MAE of TBD, MBD, and MTP contributions to intermolecular energies
      of binding for various geometries containing a grand total of over 1,300
      dimers.  All values are in kcal/mol.}
    \label{tab:mae}
  \end{center}
\end{table}

To further analyze the quality of our approach, we decomposed the MAEs of the
S22x5, S66x8, IonicHb, and X40x10 datasets at each distance factor that is
used to scale their nonequilibrium geometries (i.e.~at 0.90, 1.00, 1.25, 1.50,
2.00).  This decomposition is shown for both TBD+MTP and MBD+MTP in
Fig.~\ref{fig:dist}.  For MBD+MTP, we find the deviation from reference
systematically decaying with the intermolecular distance for all the four
datasets.  This underscores the fact that the correct dissociative
asymptotics has been taken into account via the MBD.  For smaller distances,
however, the lack of Pauli- and Coulomb-repulsion in our model appears to be
at the origin of an increasing error.  For the TBD+MTP data, however, no such
physically correct behavior is found for larger distances.  In the case of S22
and S66 the error at twice the equilibrium distance ($\sim$ 7~\AA) is in fact
as large as the error at 0.9 of the equilibrium distance.  This finding
conflicts with the common assumption that the simple $C_6/R^6$ correction will
yield proper dissociative tails.  While hardly relevant for dimers in
gas-phase, one can expect this effect to become very significant in crowded or
condensed-phase scenarios where second neighbor solvation shells contribute
significantly to the cohesive energy.  We note that the low errors for TBD+MTP
in the IonicHb and X40x10 datasets are the result of a relatively small weight
of the TBD interactions compared to electrostatics.  As such, the small error
at large distances is mainly associated to the electrostatics that has well
been accounted for through MTP.

\subsection{Water clusters}

\begin{figure}[htbp]
  \begin{center}
    \includegraphics[width=0.8\linewidth]{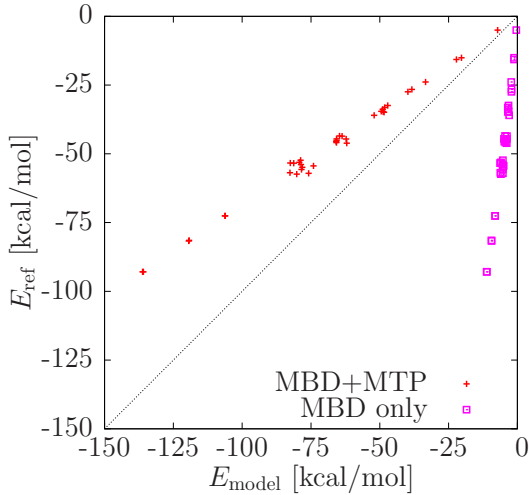}
    \caption{Predicted interaction energies from the current implementation,
      $E_{\rm model}$ = MBD + MTP versus reference energies, $E_{\rm ref}$,
      for 38 water clusters containing 2 up to 10 monomers, respectively.  The
      ``MBD only'' dataset displays the MBD component alone: Most of the
      stabilization energy arises from electrostatics.}
    \label{fig:watclusters}
  \end{center}
\end{figure}

Moving toward more complex systems, we further benchmarked the present method
against a series of 38 water clusters containing 2 to 10 water
molecules.\cite{temelso2011benchmark} The importance of many-body dispersion
in small water clusters has been highlighted by Gillan et
al.\cite{gillan2013energy} Santra et al.\cite{santra2007accuracy,
  santra2008accuracy} studied similar water clusters and reported errors that
were within 1~kcal/mol of reference \ai data using PBE0.  The results of our
approach, shown in Fig.~\ref{fig:watclusters}, indicate a strong correlation
between MBD+MTP energies versus reference CCSD(T) \ai data over a large energy
range.  As the cluster increases in size, the estimates become
overstabilizing.  For all clusters but the smallest ones, we consistently find
an error of $\approx 4$~kcal/mol per water molecule.  The MBD component is
comparatively weak, as shown in Fig.~\ref{fig:watclusters}, and illustrates
that the MTP interaction \emph{alone} overstabilizes the complex.  Defusco et
al.~highlighted the importance of repulsion to accurately describe
water-cluster geometries.\cite{defusco2007comparison} Polarizable
electrostatics is also known to play an important
role,\cite{sommerfeld2005quantum, sommerfeld2006electron} though it is likely
to lower the binding energy even further.\cite{ren2003polarizable}
Interestingly, our MTP coefficients are in good agreement with the original
AMOEBA water model.\cite{ren2003polarizable} Their excellent results in
reproducing small water clusters further suggests the role of repulsion in
this discrepancy.

\subsection{Benzene crystal}

\begin{figure}[htbp]
  \begin{center}
    \includegraphics[width=0.8\linewidth]{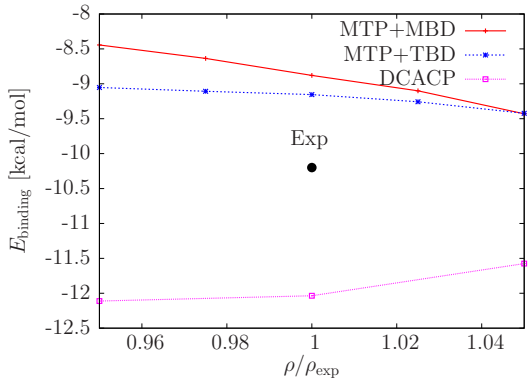}
    \caption{Cohesive binding energy of the benzene crystal as a function of
      the ratio of densities, $\rho/\rho_{\rm exp}$.  The experimental value
      is shown explicitly (black dot).\cite{schweizer2006quantum}
      The DCACP data corresponds to the BLYP + DCACP-CCSD(T) calculations of
      Tapavicza et al.\cite{tapavicza2007weakly}}
    \label{fig:bzncryst}
  \end{center}
\end{figure}

To further assess the performance of our approach, we computed the cohesive
binding energy of the benzene crystal.  Crystal structure prediction of
organic compounds has benefitted from steady progress, owing in part to the
development of both fast and accurate modeling methods (e.g.,
dispersion-corrected DFT),\cite{bardwell2011towards} to the point of ranking
polymorphs of molecular crystals.\cite{marom2013many} Akin to
Refs.~\citenum{Sprik-NoVdw} and \citenum{tapavicza2007weakly}, we computed the
binding energy for different ratios of $\rho/\rho_{\rm exp}$, where $\rho$
denotes the unit-cell density and the experimental value is taken from
Ref.~\citenum{schweizer2006quantum}.  Density scalings were performed
isotropically, while the monomers of the unit cell were translated relative to
the corresponding scaling of their center of mass.\cite{footnotebenzene}  As a
result, we restrained our study to small deviations, i.e., $|\rho/\rho_{\rm
  exp}| \le 5\%$.  The binding energy was computed using $E_{\rm binding} =
(E_{\rm complex} - \sum_{i=1}^N E_{{\rm monomer},i})/N$, where $N$ is the
number of monomers in the unit cell (i.e., $N=4$ in this case).
Fig.~\ref{fig:bzncryst} shows the binding energy as a function of
$\rho/\rho_{\rm exp}$.  We find very good agreement with the experimental
energy (i.e., $E_{\rm binding}^{\rm exp} = -10.2$~kcal/mol).  In most cases,
the MTP and MBD energy contributions are $25\%$ and $75\%$, respectively.
While Fig.~\ref{fig:bzncryst} indicates a monotonously decreasing curve, an
energy minimum is expected due to the rising contribution of repulsion as
$\rho$ gets larger.  The corresponding TBD results are on par (see
Fig.~\ref{fig:bzncryst}).  For the sake of comparison, we plot the results of
Tapavicza et al., who used dispersion corrected atom centered potentials
(DCACP) in combination with the BLYP density
functional.\cite{tapavicza2007weakly}

\subsection{Supramolecular complexes}

\begin{figure*}[htbp]
  \begin{center}
    \includegraphics[width=\linewidth]{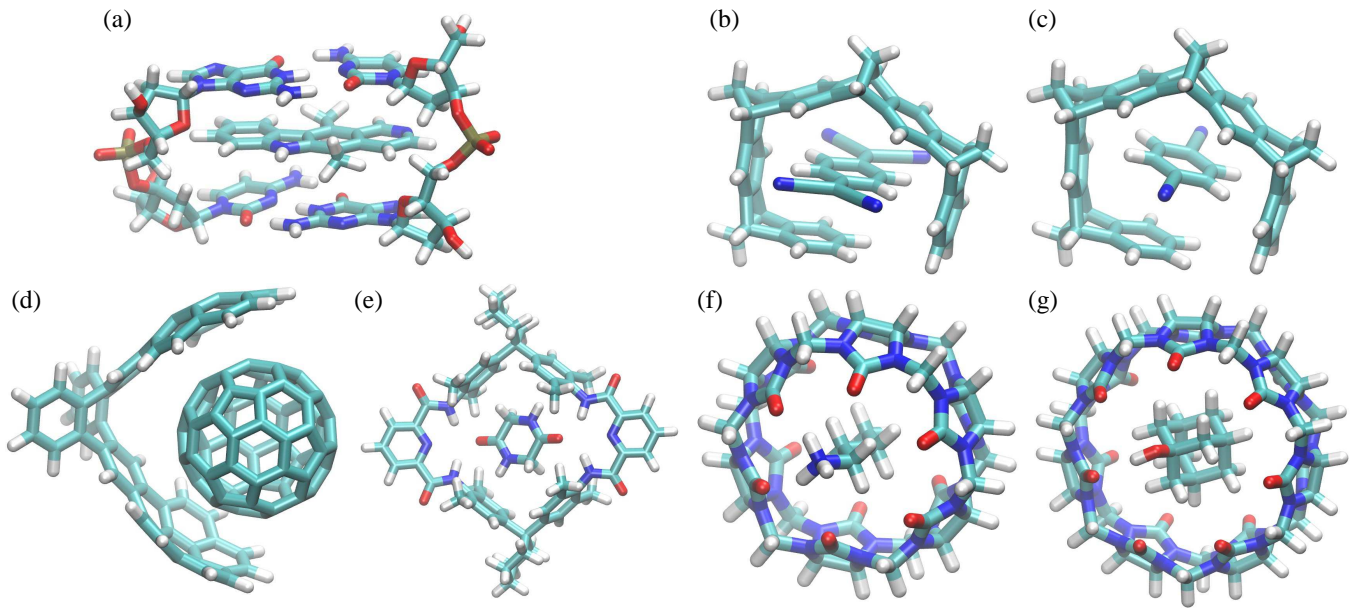}
    \caption{Cartoon representations of (a) the intercalating ellipticine
      between two Watson-Crick-bound DNA base pairs, (b)
      tetracyanoquinone-tweezer, (c) 1,4-dicyanobenzene-tweezer, (d)
      buckyball-catcher, (e) glycine anhydride-macrocycle, (f)
      butylammonium-cucurbit[6]uril cation, and (g)
      1-hydroxyadamantane-cucurbit[7]uril. Rendered in
      VMD.\cite{humphrey1996vmd}}
    \label{fig:supra}
  \end{center}
\end{figure*}

As the final and most challenging test-case, we considered several
supra-molecular host-guest complexes where one moiety encloses to a very large
degree the structure of the other moiety.  First, we calculated the
interaction energy for two Watson-Crick-bound DNA base pairs, connected
through charged sugar-phosphate backbone, into which ellipticine is
intercalated (Fig~\ref{fig:supra} (a)).  The binding energy was estimated to
amount to $\approx 37$~kcal/mol from a dispersion-corrected DFT
method,\cite{lin2007predicting} 48~kcal/mol from the Tkatchenko-Scheffler
method (i.e., without subsequent SCS),\cite{von2010two} 50.7~kcal/mol from the
non-range-separated MBD method with DFT,\cite{distasio2012collective} and
$33.6\pm0.9$~kcal/mol from diffusion quantum Monte Carlo.
\cite{benaliapplication} The complex has significant $\pi$-$\pi$ stacking and
many-body effects.\cite{von2010two} We find a binding energy $E_{\rm
  binding}^{\rm MBD} = 60$~kcal/mol, where the MTP contribution is
destabilizing (i.e., $+7.4$~kcal/mol; see Tab.~\ref{tab:supra}).  We further
note that the binding energy using TBD is significantly \emph{stronger}:
$E_{\rm binding}^{\rm TBD} = 182$~kcal/mol.  These trends are in qualitative
agreement with previous results,\cite{von2010two} though they suggest that, as
already seen for larger systems above, the current MBD implementation
overestimates the binding energy as the contact area between moieties grows.

Secondly, we also studied a select set of six host-guest supramolecular
complexes (Fig.~\ref{fig:supra} (b--g)) extracted from the S12L database of
Grimme.\cite{grimme2012supramolecular, risthaus2013benchmarking} These complexes
are formed by a host molecule (e.g., ``tweezer'', ``pincers'') and a guest
(i.e., small organic molecule) stabilized by non-covalent interactions,
including vdW interactions, hydrogen-bonding, $\pi$-$\pi$ stacking, and
electrostatic attraction.  The subset of complexes selected here follows a
recent study where diffusion quantum Monte Carlo reference numbers have been
obtained:\cite{ambrosetti2014hard} tetracyanoquinone-tweezer (b),
1,4-dicyanobenzene-tweezer (c), buckyball-catcher (d), glycine
anhydride-macrocycle (e), butylammonium-cucurbit[6]uril cation (f), and
1-hydroxyadamantane-cucurbit[7]uril (g).  Because of the sheer size of these
systems (each host contained between 72 and 130 atoms), the \ai calculations
devoted to the ESP-based fitting were run with an M06-2X functional
\cite{zhao2008m06} with 6-31+G(d,p) basis set.  Unlike the aforementioned
protocol (i.e., MTP coefficients from MP2/aug-cc-pVDZ and monopole constraints
from GDMA), we fitted the MTP coefficients of the supramolecular complexes to
the \ai ESP without any monopole constraints.  The resulting MTP and MBD/TBD
energies are presented in Tab.~\ref{tab:supra}.  Also in this case, we observe
significant overstabilization of the complexes.  Interestingly, the MBD
interactions are always less attractive than their TBD counterparts.  Different
complexes show various electrostatic contributions (e.g., the buckyball catcher
includes almost none).  This finding represents considerable numerical evidence
for suggesting that a proper inclusion of repulsive interactions and induced
electrostatics becomes all the more important as we probe more condensed or
crowded environments and scenarios.  Overall, we observe similar
overstabilization as for the water clusters (above).  Given that an accurate
account of MBD and MTP contributions significantly overstabilizes intermolecular
energies, and that induced electrostatics systematically \emph{strengthens} the
binding energy, an additive force field will require enhanced repulsive
interaction terms.  Interatomic repulsion originates in the interplay of
Coulomb-repulsion of electrons-electrons and nuclei-nuclei, as well as Pauli
exchange. As such, it appears unlikely that it can be modeled through a purely
pair-wise effective potential.

\begin{table}[htbp]
  \begin{center}
    \begin{tabular*}
      {\linewidth}{@{\extracolsep{\fill}} l|ccccc|c}
         &    MTP &   TBD &   MBD & TBD+MTP & MBD+MTP & ref.\\
      \hline
      (a) &  +7.36 & -189 & -67.0 &   -182 &  -59.6 & -33.6\\
      (b) &  -13.1 & -69.8 & -36.5 & -82.9 &  -49.6 & -27.5\\
      (c) &  -6.51 & -47.1 & -26.4 & -53.6 &  -32.9 & -17.2\\
      (d) &  +0.14 & -224  & -81.5 & -224  &  -81.3 & -25.8\\ 
      (e) &  -34.6 & -68.7 & -29.6 & -103  &  -64.2 & -33.4\\
      (f) &  -63.6 & -72.5 & -32.6 & -136  &  -96.2 & -81.0\\
      (g) &  -0.01 & -142  & -54.3 & -142  &  -54.4 & -24.1\\
      \hline
      MAE&   19.2 & 83.9  &  27.1 &  97.3  & 27.9       \\
    \end{tabular*}
    \caption{Energy contributions and reference energies (from diffusion
      quantum Monte Carlo \cite{ambrosetti2014hard, benaliapplication}) for
      the ellipticine drug (a) and the six supramolecular complexes (b--g).
      The last line indicates the MAE of each field against the reference
      values.  All energies are in kcal/mol.}
    \label{tab:supra}
  \end{center}
\end{table}

\section{Conclusion}

We introduced a Voronoi partitioning scheme for the construction of
polarizabilities of atoms in molecules, effectively coarse-graining away the
electron density.  Accounting for the heterogeneity of atoms through their
free-atom van der Waals radii provides the means to partition atoms in
molecules via a simple yet effective scheme.  Its ability to reproduce
molecular polarizabilities within the many-body-dispersion
scheme,\cite{tkatchenko2012accurate} as well as interaction energies when
combined with multipole electrostatics (MTP), makes it an appealing technique
to explore a wide variety of molecular systems exhibiting non-covalent
interactions.  Specifically, the intermolecular energies of more than 1,300
dimers were benchmarked against high-level \ai data (i.e., CCSD(T) in most
cases)---the S22 dataset alone was used for parameter optimization of the
two-body and many-body dispersion energies, where the electrostatic
contribution was subtracted from the reference energy.  The method presents an
overall mean-absolute error of $\approx 2.3$~kcal/mol---roughly on par with
standard force fields (e.g., OPLS, Amber FF03).  We highlight the excellent
performance of the TBD energies, reproducing reference data with an accuracy
of 3.1~kcal/mol.  The ease of the Voronoi-type tessellation scheme provides a
simple and efficient means to compute atomic polarizabilities from
first-principles, and extract $C_6$ coefficients that inherently depend on the
geometry of the system.  However, TBD's lack of error reduction when probing
increasingly larger distances makes a strong case for the use of MBD
interactions.

Naturally, the combined effect of dispersion energies and MTP electrostatics
still lacks repulsive interactions: several dimers, the water clusters,
ellipticine with DNA base-pair dimers, and the supramolecular complexes showed
significant overstabilization, while the benzene crystal showed no minimum
energy around the experimental unit-cell density.  A na\"ive attempt at
introducing pairwise repulsive-only Lennard-Jones-type potentials (i.e.,
Weeks-Chandler-Anderson \cite{weeks1971role}) did little in improving the
agreement with reference data (not shown)---likely due to its overly
simplifying functional form.  A comparison of our results with the original
AMOEBA water model \cite{ren2003polarizable} also points to the role of
repulsion (modeled there via Halgren's 14-7 buffered potentials
\cite{halgren1992representation}), since the other major interaction term
missing is polarizable electrostatics, likely to further \emph{stabilize}
binding.  The series of exponentials used by Whitfield and Martyna for quantum
Drude oscillators also stands as a promising
candidate.\cite{whitfield2006unified}  We finally note that the distributed
MTP methodology does not describe penetration effects, though they are
unlikely to play a large role for the systems considered here.

Performance-wise, this approach yields results in between standard force
fields and DFT methods.  TBD interactions require the calculation of effective
atomic polarizabilities using the Voronoi partitioning, which scales linearly
with the number of atoms, $N$, and the eigenvalue MBD equation will scale like
$N^3$.  In comparison, while force fields scale like $N$, DFT scales between
$N^2_{e}\ln N_e$ and $N^3_e$, where $N_e$ is the number of \emph{electrons} in
the sytem.  The computation of any complex took a fraction of a second
(slightly more for several supramolecular hosts) on a single core.  It stands
to reason that a number of obvious optimizations (e.g., parallelization) could
be undertaken to reduce the computational investment.  We point out that the
present work achieves classical intermolecular energies with similar
accuracies as standard force fields, without experimental input---as achieved
earlier by other methods \cite{piquemal2007toward, gordon2013accurate}---and
negligible parametrization effort: the TBD/MBD energy has no free parameter
(excluding the transferable parameters that were set in this work)
irrespective of the chemistry at hand (i.e., no need for specific atom
types)---though heavier elements may present additional complications---while
systematic protocols can fit MTP coefficients from the \ai ESP in an automated
way.  We further note that this prior \ai calculation is not a necessary
curse: an optimized workflow could involve a fragment-based library of
precomputed MTP coefficients.  The limited computational overhead associated
with the TBD method would make its use in force fields conceivable: the
inherent dependence of the $C_6$ coefficients on the geometry and chemistry
(e.g., hybridization state) of the system may help reach more transferable
force fields---a critically lacking feature of current atomistic models, which
require extensive effort for any new (i.e., yet unparametrized) molecule
studied.  As for MBD, the eigenvalue equation involved hampers the prospects
of an efficient MD implementation.  However, it provides interesting
perspectives in the context of well-established potentials (e.g.,
SIBFA,\cite{piquemal2007anisotropic} EFP\cite{gordon2013accurate}) that aim at
computing accurate intermolecular energies for large systems.

\begin{acknowledgments}
  We thank Alexandre Tkatchenko and Denis Andrienko for insightful discussions
  that shaped the present work. We also thank Sanghamitra Neogi, Davide
  Donadio, and Markus Meuwly for comments and critical reading of the
  manuscript.  TB acknowledges access to Markus Meuwly's computational
  resources.

  TB was partly supported by the Swiss National Science Foundation through the
  NCCR-MUST and grant 200021-117810.  OAvL acknowledges support from the Swiss
  National Science foundation No.~PPOOP2\_138932.  This research used
  resources of the Argonne Leadership Computing Facility at Argonne National
  Laboratory, which is supported by the Office of Science of the U.S.~DOE
  under Contract No.~DE-AC02-06CH11357.
\end{acknowledgments}


\begin{thebibliography}{100}
\bibitem{footnotebenzene}
Note that the current implementation does not allow to relax internal
coordinates.
\end{thebibliography}
\end{document}